# The Arctic Electric Power Stations Are the Decision of Energy, Environmental and Climate Problems.

B.M.Ovchinnikov[1]*, Yu.B.Ovchinnikov[2], V.V.Parusov[1]

[1]Institute for Nuclear Research, Russian Academy of Sciences, Moscow, Russia

[2]National Physical Laboratory, Hampton Road, Teddington, Middlesex, UK

*Corresponding author: ovchin@inr.ru

## Abstract

In this paper we propose the decision of energy, environmental and climate problems of the society with the help of arctic electric power stations. Such stations can provide the electricity to the whole of Europe and Asian, this can save from dependence on oil and gas. This method of producing electricity is completely ecologically safe. The stations are decision of the energy problem in the case of "nuclear winter". The paper contains project on establishment of arctic (winter) power stations (see our publication: Journal of Thermal Engineering, Vol.51, №2, 2004, B.M. Ovchinnikov et al. "Power plants that use the thermal energy of natural water bodies and the atmosphere")[1]. The work is based on the power that is released, when water freezes with releasing of large energy (333 Joules per 1 gram of water). The general principle of the station consists in next. The water is frozen on the boiler, which is located in the water (river, lake or ocean), in the boiler the working fluid evaporates, enters the turbine and after that it is fed into the cooling tower, the cold air cooled, liquefied and pumped back to the boiler etc.

## 1. Introduction

The main problem humanity must solve in the near future is the energy problem. Although the available deposit of hydrocarbon gases under the seabed are enough to last several hundreds of years, the continent combustion of hydrocarbons fuels and the results buildup carbon dioxide in the atmosphere is unacceptable because of the growing greenhouse effect and the approaching environmental catastrophe. Nuclear power poses a risk to the environment through the possibilities of another Chernobyl-type (or Fucushima-type) disaster and the yet unresolved problem of nuclear-waste recovery. Moreover, the supply of uranium and thorium from natural deposits is limited to a few centuries.

Renewable energy sources (RESs) and power stations based on wind, solar, hydro, and geothermal energy have not such disadvantages and they do not emit carbon dioxide into the atmosphere. Recent years have seen the development of power stations using the thermal energy of natural bodies of water (seas, oceans, rivers, and lakes). These RESs and power stations are also free from the above negative effects. Compared to conventional energy sources, the advantage of RESs is their virtual inexhaustibility and environmental friendliness [2-4].

As early as 1881, French physicist Jacques-Arsene d'Arsonval first demonstrated the possibly of developing a heat engine that could operate on a temperature difference between a heater and a cooler of about 20°C. Reference [5] describes using the energy of naturally heated water in warm zones of the World Ocean for industrial power production. Even more efficient were the designs for floating large-capacity (100-MV) OTPS systems developed by the American "Lockheed" and SSP companies in 1978 and 1981, respectively [6].

Arctic electric power stations that use the difference between the temperatures of the sub-ice sheet waters and of the air are being developed [7].

## 2. Operation principle of arctic electric power station.

The work of arctic station is based on the power that released, when water freezes (1 gram of water releases 333 Joules of heat). The station contains the working body (gas), which is liquefied at the arctic temperature (about minus 30 0°C) and evaporates at the temperature about 0°C. The station operates on organic Rankine cycle. The station consists of turbine, electric generator, pumps, condenser, cooling tower, boiler-evaporator and buffer (Fig.1).

The boiler is locates in water (river, lake or ocean). The working body is liquefied in the cooling tower, and pumped to the boiler, the water is frozen on the surface of the boiler with the heat releasing, the working body is evaporates, enters the turbine, makes the work and then enters to the cooling tower, liquefied and after that is pumped to the boiler etc. The working fluid of the unit is a volatile liquid with a low freezing point. The evaporator, the turbine and the condenser contain the working fluid at pressures than enable the fluid to evaporate at the temperature that exists in the evaporator and condense at the temperature attained in the condenser. In fig. 2-4 we present the design of cooling tower, condenser and boiler-evaporator. The turbine, generator, pumps and buffer are standard industrial devices.

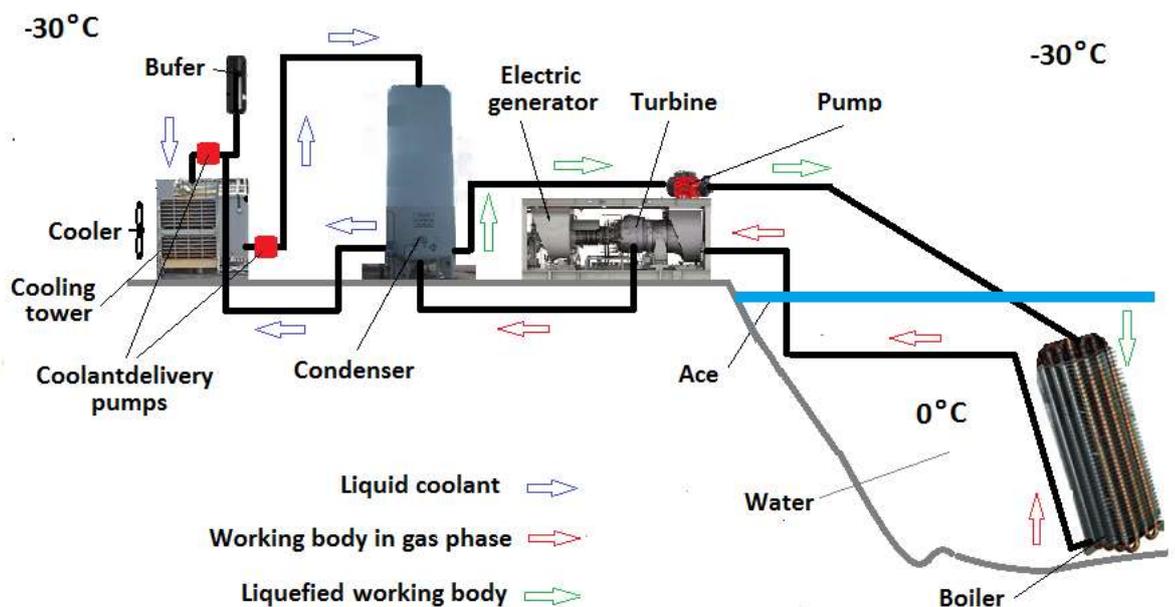

Fig.1 Operation principle of power station

## 2.1 Construction of boiler-evaporator.

Figure 2 shows the design of a tubular boiler-evaporator. It is installed in sub-ice sheet water having a temperature close 0°C. The evaporator tubes are quite closely oriented to the vertical position. At the same time, the distribution of the working fluid allows the liquid to flow down along the tubes in uniform layers. Such a regime ensures maximum heat transfer from the tube walls to the working fluid.

The water temperature is close to zero, and the working fluid absorbs heat from the water, heats up, and evaporates; therefore, a certain icing of the tube surfaces during operation is inevitable. The heat release is 333 Joules per gram of ice, that is, intense heat transfer from the freezing water to the tube walls takes place.

This method of heating the evaporator tubes by icing them is much more efficient then one based on supplying a great deal of water with a near-zero temperature to the evaporator [7].

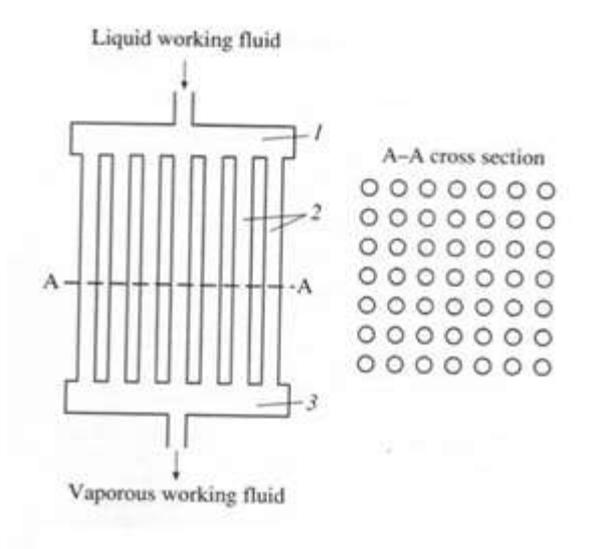

Fig.2 Boiler-evaporator

1-distributor of the working fluid, 2-system of vertical parallel evaporator tubes, 3-vapor collector

## 2.2 Construction of condenser.

Figure 3 shows the design of a shell-and-tube condenser. The exhausted vaporous working fluid enters the system of inclined ribbed parallel tubes immersed in the coolant. The working fluid condenses in them, pump delivers the liquefied working fluid to evaporator. From the condenser, the coolant is pumped to the cooling tower. In the latter, the coolant is distributed among plates that are cooled with cold air from the fan, and flows down along these plats as a thin layer. The cooling tower surpasses the direct cooling of the condenser tubes with could air because of the following:

First, according to [13], the heat transfer coefficient between a thin layer of the coolant and the forced air flow in the cooling tower is an order of magnitude greater then that of air blowing over the condenser tubes;

Second, if necessary, the effective heat transfer surface area can be made much larger in a cooling tower than the corresponding surface area in a tubular exchanger.

The cooling tower may use as coolant, for instance, ethanol, which has a freezing point of -114°C, or a calcium-chloride solution as a coolant [7].

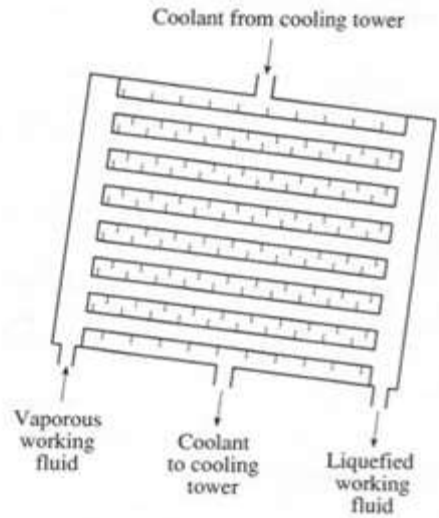

Fig.3 Condenser

## 2.3 Construction of cooling-tower.

Figure 4 shows the design of the cooling-tower. The latter is a system of parallel plates channeling down the thin layers of the coolant. The fan blows over the plates with the cold ambient air. In such a cooling tower, the coefficient of the heat transfer can reach 250 W/(m² K), that is, an order of magnitude greater than that in a shell-and-tube heat exchanger. With an air temperature of -40°C and ΔT≈5°C, we have the heat flux density N≈1,25 kW/m². The removal of 1mW of heat in the condenser requires the plates to have a total surface area of 800 m². The cooling tower comprises 80 plates, each having a surface area of 10 m².

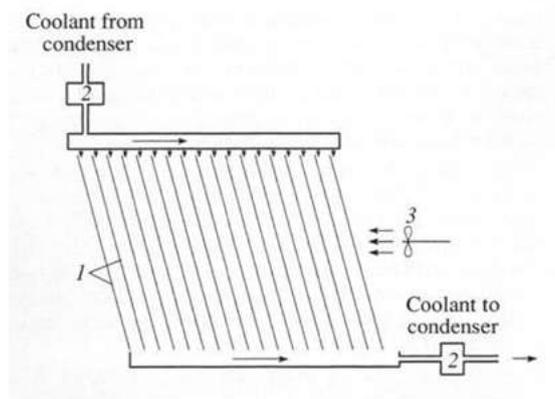

Fig.4 Cooling-tower

1-cooling-tower plates, 2-coolantdelivery pump, 3-cooler.

## 3. Coefficient efficiency of arctic electric power stations

Let us assess the efficient of the choice of the heat engine from the following formula derived from the law of conservation of energy [8-10]:

$$\eta = (1 - q_c/q_e) \chi = \eta_0 \chi,$$

Where $q_c$ is the heat absorbed by the working fluid the boiler; $q_e$ is the heat removed from the working fluid in the condenser; $\chi \approx 0{,}75$ is the efficiency of the turbine; and $\eta_0$ is efficiency of cycle, ignoring the losses in the turbine.

The following formulas determine the values of q [1].

The work $A_u = q_c - q_e$ and the thermal efficiency $\eta_0$ can be determined from the tabulated data of the thermo-physical properties of gases and liquids in [8, 10, 11]. Table [11] contains the result of calculations including the $A_u$ for a reversible Rankine cycle and the values of $\eta_0$ for several working fluids [12].

The determined values $\eta_0$ show that virtually all of working fluids have suitable characteristics for turbine operation. For the winter temperature difference of 20°C and more, the efficiency of the reversible cycle exceeds a value of approximately 0,07, with the dryness of the vapor in the turbine being $\chi > 0{,}88$. For instance, the efficiency of the heat engine working with propane at T=268 K and T=231 K is 0,095, while the efficiency of the heat engine working with ethylene at T=268 K and T=228 K is 0,1095.

The need to search for inflammable working fluids that have better properties than propane and ethylene should be stressed.

## 4. Ecologically safety

The method is completely ecologically safe, in contrast to gas-, oil- and coal-stations. The arctic power station has no problems with waste disposal. Oceanic (summer) power stations [5-7] operating on the Rankine cycle of open-cycle type lift cold water (4 0°-6 0°C) from a depth of about 1 km to the surface of the ocean with an average temperature of 24 0°-26 0°C at the equator zone. Water pressure is relieved and a large amount of carbon dioxide ($CO_2$), which in oceanic water is released. Oceanic power stations emit 1.5 times more $CO_2$ per 1 MW of generated power than hydrocarbon-fueled stations. Arctic electric power stations have not such disadvantages and they do not emit $CO_2$ into the atmosphere. The elements of power stations based on wind energy (electric battery) and solar energy (solar elements) are not environmentally friendly in terms of their production and disposal [4].

## 5. Mobile and floating arctic electric power stations

At present, electricity supply program is being implemented in the Russian remote northern territories with the help of floating nuclear power plants, but the program is not fully solved the problem of disposal of spent nuclear fuel.

We propose to solve the problem of supplying these areas with the help of mobile arctic plants. This station must consist of a container which contains the turbine, generator, pump, the cooling tower, condenser and boiler-evaporator unit blocks are connected with releasable hoses to turbine. Such stations can be transported by helicopter. It is possible to use floating arctic power plants similar to the above-mentioned floating nuclear power plants.

## 6. Nuclear Winter

These power plants solve the energy problem in the event of a "nuclear winter" (a sharp drop in temperature on the ground in the event of a massive meteorite falling, or a super volcano or nuclear war. In the case of a "nuclear winter", it is impossible to grow crops in the fields, and the only way to grow the crop will be a hothouse, requiring a lot of energy, but the reserves of nuclear and hydrocarbon fuels on earth are limited, so only arctic (winter) power stations will provide mankind with food.

## 7. Green hydrogen generation

Climate change is a topic that has been actively discussed in the widest circles in recent years, including in the political arena. To solve the problems of climate conservation, 13 leading energy, transport and industrial companies - ENGIE, Royal Dutch Shell, Total, Alstom, Linde Group, Toyota, BMW GROUP, Daimler, Honda, Hyundai Motor, Kawasaki, Air Liquide, Anglo American - have joined forces to launch a new hydrogen-based energy model. To this end, the Hydrogen Council was established in Davos on January 17, 2017 [14].

Currently, the authors are developing an optimal scheme for generating green hydrogen at the arctic power plants by electrolysis of water (energy consumption of 5 KW/h per 1 m³ of hydrogen) or adiabatic conversion of methane in hydrogen (energy consumption of 1 KW/h per 1 m³ of hydrogen) [15].

As well as, we are developing energy system for storing and transporting green hydrogen produced at the hydrolyser (Fig.5 right) or adiabatic convertor (Fig.5 left). It is possible to transport in Europe of green hydrogen (or hydrogen-methane mixture) through existing northern gas pipelines.

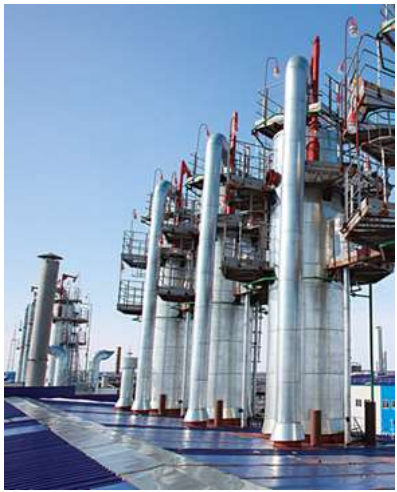
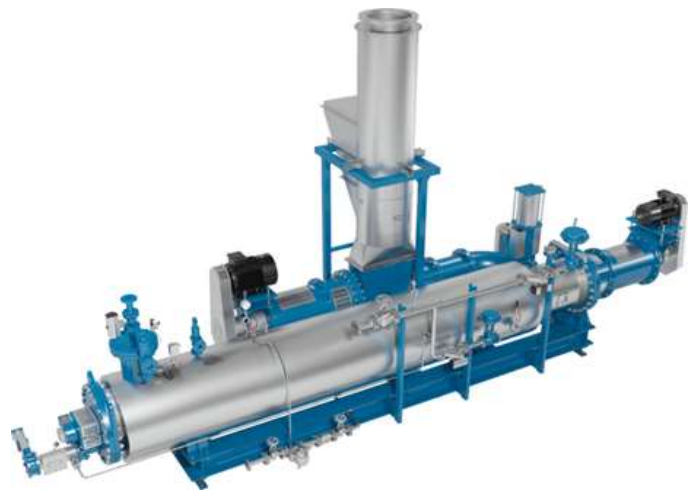

Fig.5 Adiabatic convertor of methane in hydrogen (left) and hydrolyzer of water (right)

## 8. Conclusion

Global warming and climate change mitigation are the present most challenging task of humanity. The generation of only one MWh of energy emits up to 1000 kg of transparent $CO_2$ into the atmosphere.

It is necessary to develop the arctic power station for decision of climate problem of mankind, in case of "nuclear winter" and for decision of energy and environmental problems. Many countries have a huge sea-cost on the Arctic Ocean. Such powers can provide the electricity to the whole of Europe and Asia. This can save from dependence on oil and gas*. Green hydrogen can be produced via electrolysis using arctic station electricity. We invite to cooperation for construction such electric power station.

*The station was nominated on Global Energy Prize in 2018-2020. The Global Energy Prize annually honors outstanding achievements in energy research and technology from around the world that are helping address the world's various and pressing energy challenges. The Global Energy Prize, founded in 2002, is awarded to the most accomplished minds in the research world. The honorees are awarded the Global Energy Prize at the International Economic Forum in St Petersburg along with a prize of RUB 39 million.